\documentclass[aps,prb,preprint,superscriptaddress,showpacs]{revtex4-1}

\usepackage{graphicx,psfrag}
\usepackage[latin1]{inputenc}
\usepackage{natbib}
\usepackage{amsmath, amsthm, amssymb}
\usepackage{color}
\usepackage{soul}
\usepackage{comment}
\usepackage{hyperref}
\usepackage{lineno}

\newcommand{\notes}[1]{}

\newcommand{\beq}{\begin{equation}}
\newcommand{\eeq}{\end{equation}}
\newcommand{\beqnn}{\begin{equation*}}
\newcommand{\eeqnn}{\end{equation*}}
\newcommand{\beqas}{\begin{eqnarray*}}
	\newcommand{\eeqas}{\end{eqnarray*}}
\newcommand{\beqa}{\begin{eqnarray}}
\newcommand{\eeqa}{\end{eqnarray}}

\begin{document}


\title{Generation and imaging of magnetoacoustic waves over millimetre distances} 

\author{Blai Casals}
\affiliation{Institut de Ci\`encia de Materials de Barcelona (ICMAB-CSIC),Campus UAB, 08193 Bellaterra, Spain}
\author{Nahuel Statuto}
\affiliation{Dept.\ of Condensed Matter Physics, University of Barcelona, 08028 Barcelona, Spain}
\author{Michael Foerster}
\affiliation{ALBA Synchrotron Light Source, 08290 Cerdanyola del Vall\`es, Spain}
\author{Alberto Hern\'andez-M\'inguez}
\affiliation{Paul-Drude-Institut f\"ur Festk\"orperelektronik, Hausvogteiplatz 5-7, 10117 Berlin, Germany}
\author{Rafael Cichelero}
\affiliation{Institut de Ci\`encia de Materials de Barcelona (ICMAB-CSIC),Campus UAB, 08193 Bellaterra, Spain}
\author{Peter Manshausen}
\affiliation{Institut de Ci\`encia de Materials de Barcelona (ICMAB-CSIC),Campus UAB, 08193 Bellaterra, Spain}
\author{Ania Mandziak}
\affiliation{ALBA Synchrotron Light Source, 08290 Cerdanyola del Vall\`es, Spain}
\affiliation{Instituto de Química Física "Rocasolano", Madrid, 28006, Spain}
\author{Lucía Aballe}
\affiliation{ALBA Synchrotron Light Source, 08290 Cerdanyola del Vall\`es, Spain}
\author{Joan Manel Hern\`andez}
\affiliation{Dept.\ of Condensed Matter Physics, University of Barcelona, 08028 Barcelona, Spain}
\author{Ferran Maci\`a}
\affiliation{Institut de Ci\`encia de Materials de Barcelona (ICMAB-CSIC),Campus UAB, 08193 Bellaterra, Spain}
\affiliation{Dept.\ of Condensed Matter Physics, University of Barcelona, 08028 Barcelona, Spain}
\email{ferran.macia@ub.edu}

\date{\today}

\begin{abstract}
Using hybrid piezoelectric/magnetic systems we have generated large amplitude magnetization waves mediated by magneto-elasticity with up to 25 degrees variation in the magnetization orientation. We present direct imaging and quantification of both standing and propagating acoustomagnetic waves with different wavelengths, over large distances up to several millimeters in a nickel thin film.
	
\end{abstract}

\maketitle 

Magnetization oscillations in thin films with spatial variations at the nano/micro scale are interesting in the developing devices for high-speed and low-power signal processing compatible with existing technology. In particular, spin waves are collective excitations of magnetic order in materials having frequencies and wavelengths determined by the strength of magnetic interactions---mostly dipolar and exchange interactions. In ferromagnets the typical frequencies are in the low GHz regime with wavelengths from hundreds of nanometers---dominated by quantum exchange---to a few micrometers---dominated by dipolar fields. Collective magnetization oscillations are traditionally excited via spatially non-uniform oscillating magnetic fields generated by antennas or strip lines with an electrical current \cite{Vlaminck410}. However, generation of magnetization oscillations with high amplitudes and over long distances is challenging due to the mismatch of wavelengths with electromagnetic waves in free space, which is of the order of several centimeters. The spin-transfer-torque effect \cite{Ralph2008,Brataas2012} generated from spin currents or ultrashort laser pulses \cite{Mangin2014,Bayer2017} provide promising alternative pathways towards the control of dynamic magnetic states at the nanoscale without using electromagnetic fields. In metallic ferromagnets the magnetic damping, however, results in low amplitude and/or strongly localized excitations, limiting the applicability in devices. Using hybrid piezoelectric/magnetic systems we have generated large amplitude magnetization waves mediated by \textit{magneto-elasticity} with up to 25 degrees variation in the magnetization orientation. We present direct imaging and quantification of both standing and propagating magnetization oscillations with different wavelengths, over large distances up to several millimeters.


A promising strategy for handling magnetization variation at the nanoscale together with low-power dissipation is the use of strain. The magnetoelastic effect (ME) is the change of  material's magnetic properties under an elastic mechanical deformation---strain. A change in atomic distances caused by strain modifies magnetic interactions, resulting in a ME-induced anisotropy. Surface acoustic waves (SAW) are strain waves that may propagate millimetre distances at the surface of a material and can be generated with oscillating electric fields in a piezoelectric material \cite{SAW1,SAW2}. Today, radio-frequency filters and delay lines based on SAW are a standard technology used in mobile phones because of their ability to convert a centimetre wavelength in free space into a micrometer wavelength in a chip. 

Recent interest has focused on the interaction between electrically generated acoustic modulations and magnetization dynamics. Clear observation of the interactions such as changes in SAW propagation caused by the back action of the magnetization dynamics \cite{Weiler2011, Ralph2015,Labanowski2016} or the variation of magnetic states caused by SAW \cite{Hernandez2006, Davis2010,Weiler2012, Thevenard2016, NatCom2017_Foerster, Kuszewski_2018,MRS2018, Labanowskieaat6574}, including observations of the oscillatory magnetic signal associated to a magnetoacoustic wave \cite{janusonis_2015_apl,thevenard2018_pra,Thevenard_2019, Carman_2014,Optical_pump_SAW} have been reported. There exist also the possibility to generate spin currents via spin-rotation coupling using a surface acoustic wave \cite{Kobayashi}. In addition, some theoretical studies suggest that spin waves coupled to strain waves may be amplified and propagate longer distances \cite{Duflou2017,Carman_sim_2017,slahuddin_SAW_spin,paramametric_prb2019}. 

In order to understand the interaction between acoustic waves and magnetization waves and to determine the feasibility of generating magnetization oscillations with SAW we designed an experiment to resolve the evolution of both waves in space and time. We fabricated hybrid devices of piezoelectric (LiNbO$_3$) and ferromagnetic (nickel) materials with interdigitated (IDT) antennas capable to generate SAW in the range of 0.1 to 2.5 GHz. Our measuring technique combines time and spatially resolved Photoemission Electron Microscopy (PEEM) to obtain electrical contrast of the SAW with X-ray magnetic circular dichroism (XMCD) \cite{Stohr2007, Aballe2015} to achieve magnetic contrast of magnetization waves. The SAW excitations are synchronized to the repetition rate of the ALBA synchrotron and stroboscopic images of strain and magnetization are obtained simultaneously \cite{Foerster2016,NatCom2017_Foerster}. We show that acoustic waves determine the spin orientation in the ferromagnet and we obtain a large variation---up to 25 degrees---of the magnetization rotation between opposite phases of the wave at room temperature. It is, thus, possible to generate large-amplitude magnetization waves mediated by strain---\emph{magnetoacoustic waves}---with wavelengths of a few micrometers propagating over millimetre distances in a thin ferromagnet. Different frequencies and wavelengths can be achieved in the same device and the amplitude of magnetoacoustic waves can be modulated with an external applied magnetic field. Combining multiple SAW excitations we can also generate interference patterns of magnetoacoustic waves. 

\begin{figure}[htb!]
	\includegraphics[width=1\columnwidth]{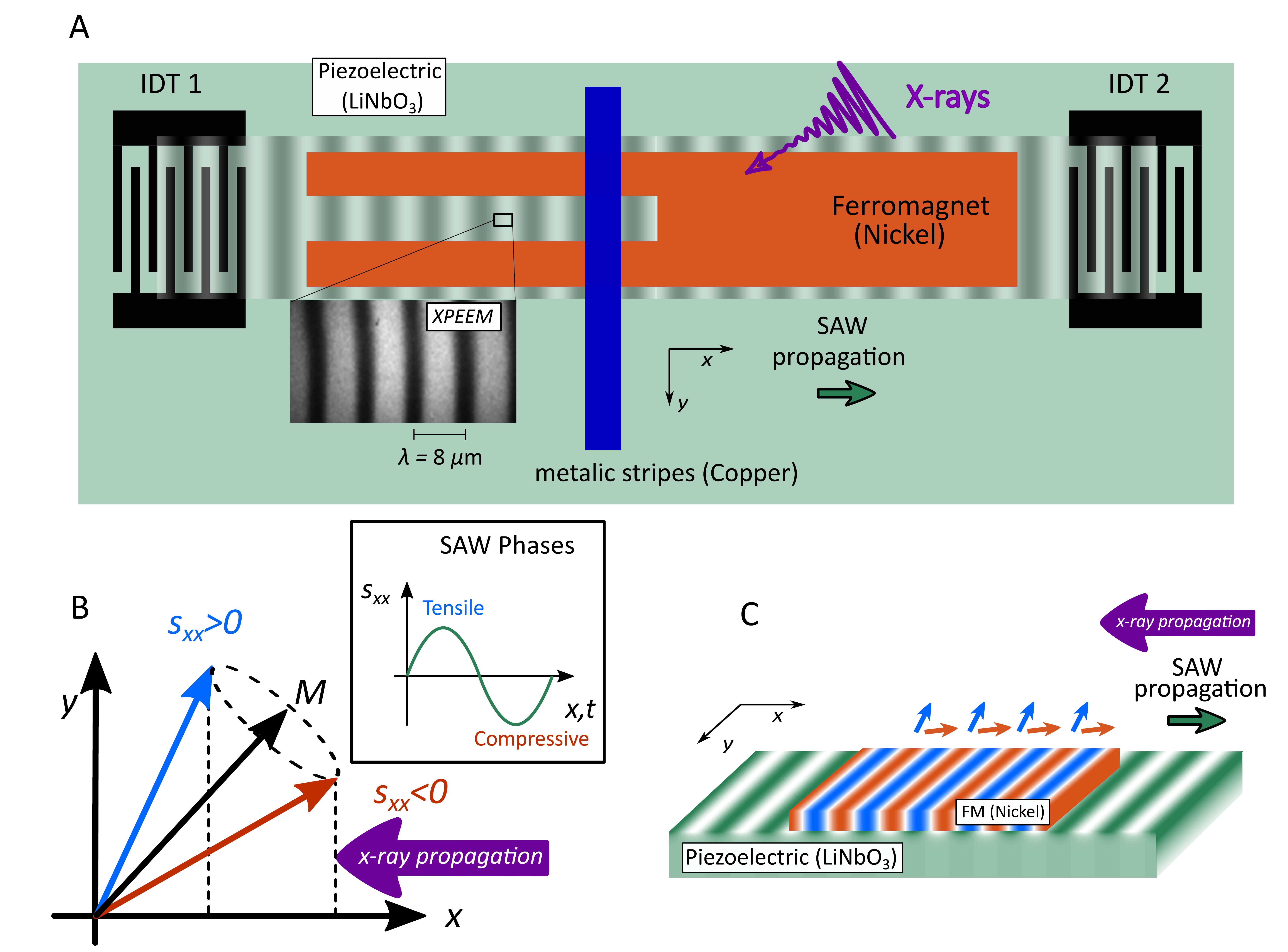}
	\caption{\small{\textbf{(A)} Schematic set up of the hybrid devices with ferromagnetic U-shape structures (nickel) on a piezoelectric substrate (LiNbO$_3$), The structures are deposited on the acoustic path between two IDTs. Additional non-magnetic Copper structures are also included in the hybrid devices. The X-rays illuminate the sample with short pulses of 20 ps and a frequency of 500 MHz. The interdigital transducer, IDT1, is excited with an AC electric signal which generates a surface acoustic wave (SAW) propagating through the LiNbO$_3$. An image at the center of the acoustic path taken with a PEEM microscope is shown as an inset. Vertical stripes result from the piezoelectric voltage associated to the SAW.  \textbf{(B)} Schematic representation of the magnetization oscillation as a function of the SAW strain phases. \textbf{(C)} Cartoon of the in-plane strain caused by the SAW in the piezoelectric (in green colorscale) and magnetic modulation in the ferromagnet (in orange-cyan colorscale).}}
	\label{fig1}
\end{figure}

First, we characterize the strength of SAW by direct XPEEM images of the piezoelectric substrate LiNbO$_3$. Within the acoustic path, as shown in Fig. \ref{fig1}A, bright and dark stripes with the periodicity of the SAW excitation ($\lambda\sim$ 8 $\mu$m) are visible, independently of the X-ray photon energy or polarization. This contrast in the energy-filtered PEEM images is due to the fact that the piezoelectric voltage at the surface associated with the wave shifts the energy of secondary electrons leaving the sample surface. This effect is only visible on the  LiNbO$_3$ substrate surface; any metallic structure on the LiNbO$_3$ fully screens the electric field and prevents the SAW observation. The strain amplitude produced by the SAW can be calculated from the value of the piezoelectric surface potential obtained from XPEEM images \cite{Foerster_JSR}. In our experiment we apply a maximum SAW power corresponding to 0.05\% of strain. We assume the displacement oscillation along $x$, the propagation direction of the SAW, to be fully transferred to the nickel thin film---which is 20 nm thick whereas the SAW penetration depth is of the order of micrometers.

The signal intensity in the XMCD images---the difference between images obtained with opposite circular polarization of the incident x-rays---is proportional to the nickel magnetization component along the X-ray incidence direction, which is aligned with the SAW propagation (See Fig. \ref{fig1}AB). The strain associated to the SAW induces a varying magnetic anisotropy in the nickel film that oscillates between the SAW propagation direction and its orthogonal direction. We apply a small magnetic field, $\sim$1 mT, with an angle of $\sim$60 degrees with respect to the SAW propagation direction in order to maximize the effect of the induced magnetoelastic anisotropies (See Supplementary Text). Figure \ref{fig2}A shows PEEM images of the magnetic contrast of the nickel film taken at the nickel $L_3$ absorption edge energy under a SAW excitation of 500 MHz. The XMCD image of Fig. \ref{fig2}A  shows a wave pattern in the ferromagnetic film and nothing in the piezoelectric film (the PEEM signal in the piezoelectric does not depend on the x-ray helicity). The PEEM image of Fig. \ref{fig2}B corresponds to a subtraction of two images with equal x-ray helicity but with opposite phases in the SAW excitation (180 degrees difference): the result is that the dynamic magnetic contrast in the nickel film caused by the SAW is as clear as in XMCD images but now we also detect the surface potential in the piezoelectric. We note that XMCD images capture both static and dynamic variations of magnetization whereas PEEM images with a 180-degree phase subtraction provide only information on variations produced by the SAW.

Figures \ref{fig2}A and \ref{fig2}B show a magnetization wave of 500 MHz with 8 $\mu$m wavelength. We measure a variation of the XMCD signal up to 0.02 in the magnetic wave whereas the variation between opposite magnetised sample ($\pm M_s$) is 0.1; we estimate thus an amplitude of the magnetoacoustic waves of about 40\% of $M_s$, which corresponds to about 25 degrees (peak to peak). Figure \ref{fig2}C shows traces of the magnetoacoustic waves at different applied SAW amplitude and Fig. \ref{fig2}D summarizes the dependence of the magnetic wave amplitude with strain, which in the studied regime is linear. The strain value associated to the SAW is calculated from the photoelectron shift of the local amplitude of surface potential\cite{Foerster_JSR}. 

The observed inhomogeneous magnetization oscillation are in the long wavelength limit where the frequency has a weak dependence on the wavelength---exchange interaction hardly plays a role here. However, the dispersion relation---frequency that corresponds to each wavelength---for the SAW follows the expression $f=v\lambda$, where $v$ is the speed of sound in the LiNbO$_3$ and $\lambda=\lambda_0/n$ is the wavelength with $\lambda_0$ determined by the IDT fingers periodicity and $n$ the excitation order. To study the wavelength dependence of the magnetoacoustic waves we use a gating system in the microscope that allows synchronizing at submultiple frequencies of the synchrotron repetition rate \cite{ABALLE201910}. Figures \ref{fig2}E-H show magnetoacoustic waves of frequencies 125, 250, 375, and 500 MHz with wavelength of 32, 16, 12, and 8 $\mu$m, in all cases reaching a magnetization oscillation of several degrees between opposite phases.

A magnetic field of about 3 mT is enough to reverse the magnetization of the nickel layer, which has a small uniaxial anisotropy ($k_U \simeq 800 $J m $^{-3}$) in the $x$ direction caused by the deposition process. Samples include a non-magnetic Copper structure in addition to the ferromagnetic film in order to provide a zero level reference of the XMCD signal. Figures \ref{fig3}A-C show consecutive XMCD images at applied fields before, during, and after magnetization reversal in presence of a SAW excitation. In Fig. \ref{fig3}A magnetoacoustic waves superimposed to an homogeneous magnetization pointing opposite to the magnetic field (in red) are visible within the ferromagnet. In Fig. \ref{fig3}B the magnetization begins to reverse forming domains (in blue) with the magnetization pointing towards the applied magnetic field; magnetoacoustic waves are visible in both magnetic domains. Finally in Fig. \ref{fig3}C a slightly larger applied field reverses all film's magnetization and magnetoacoustic waves are still visible over the whole nickel film. Figure \ref{fig3}D shows horizontal cuts from Fig. \ref{fig3}A and \ref{fig3}C together with a schematic arrow plot indicating the magnetization direction at different wave points. An additional horizontal plot corresponding to a larger applied field of 9.5 mT is also plotted in the same graph in order to show the reduction of the magnetization oscillation amplitude.

A summary of the effect of a magnetic field is shown in Fig. \ref{fig3}E. The amplitude of magnetoacoustic waves as a function of the applied magnetic field shows a peak at a small field of about 5 mT---and vanishes at larger magnetic fields. The magnetic field is applied in the film plane with an angle of $\sim 60$ deg. with respect to the SAW propagation direction in order to maximize the effect of the SAW induced anisotropy onto the magnetization variation \cite{Weiler2011}. Additional field angles, including a magnetic field applied in the SAW propagation direction and showing no effect on the measured magnetization, were also explored and are presented in the Supplementary Text. 


The magnetic torque produced by strain on the ferromagnet competes with the applied field torque and that of the growth-induced uniaxial anisotropy. At large applied fields the effect of strain on the magnetization is negligible, and the latter aligns with the applied field; at zero applied field the dominating contribution is the uniaxial anisotropy along the $x$ direction. The effect of strain on the magnetization is maximal when the applied field is comparable to that caused by the uniaxial anisotropy (around 3 mT). This argument holds for a ferromagnet with an intrinsic anisotropy comparable to the one induced by the SAW and disregards resonance effects \cite{Weiler2011,Ralph2015,Labanowski2016,Labanowskieaat6574}. Our micromagnetic simulation account for all the mentioned effects \cite{Kuszewski_2018} including exchange and dipolar interactions and provide us with an estimation of the intrinsic anisotropy value of $\sim$800 J/m$^3$, and the SAW induced anisotropy of $\pm$600 J/m$^3$, which matches with earlier measurements, \cite{NatCom2017_Foerster}. 


It is also possible to create and control standing SAW \cite{Beil2008, Lima2012,Foerster_JSR} and, thus, to stabilize spatial strain patterns (alternating nodes and antinodes) over any desired period of time (much longer than the SAW period). Using two IDTs facing each other we create and control patterns of standing magnetoacoustic waves. Videos of PEEM images where opposite phases have been subtracted in order to obtain magnetic contrast in the ferromagnetic film can be seen in the Supplementary Material for both propagating and standing magnetoacoustic waves.  Figure \ref{fig4} shows the evolution of a magnetoacoustic waves in time both for propagating, Fig. \ref{fig4}A, and standing waves, Fig. \ref{fig4}B, obtained from the videos.

The present results demonstrate an alternative and technologically feasible technique to generate magnetization waves of large amplitude, tens of degrees, over long distances, up to centimetres, i.e., several orders of magnitude longer than Oersted-field driven spin waves. Besides the fundamental interest of the creation and observation of magnetoacoustic waves---including standing patterns---these findings may open new directions in fields requiring strong and localised dynamic magnetic fields \cite{Labanowskieaat6574}. Further work might explore the possibilities of having carrier signals on the propagating magnetoacoustic waves or of using the extended--and time varying---inhomogeneous magnetization pattern.

\newpage

%

\section*{Acknowledgments}

Authors acknowledge Jordi Prat of ALBA for his help during experiments. FM. acknowledges support from the MINECO through Grant No. RYC-2014-16515. FM, BC, and RC acknowledge support from MINECO through Grants No. SEV-2015-0496 and No. MAT2017-85232-R. FM, JMH and NS acknowledge funding from MINECO through Grant No. MAT2015-69144-P. We thank Werner Seidel from PDI for assistance in the preparation of acoustic delay lines on LiNbO$_3$. LA and MF acknowledge support from MINECO through RTI2018-095303-B-C53. This project was partially supported by the ALBA in-house research program through projects ALBA-IH2015PEEM and ALBAIH2017PEEM.


\section*{Supplementary materials}
See Supplementary Text\\
See Suplementary movie


\clearpage

\section*{Figures}

\begin{figure}[htb!]
	\includegraphics[width=\columnwidth]{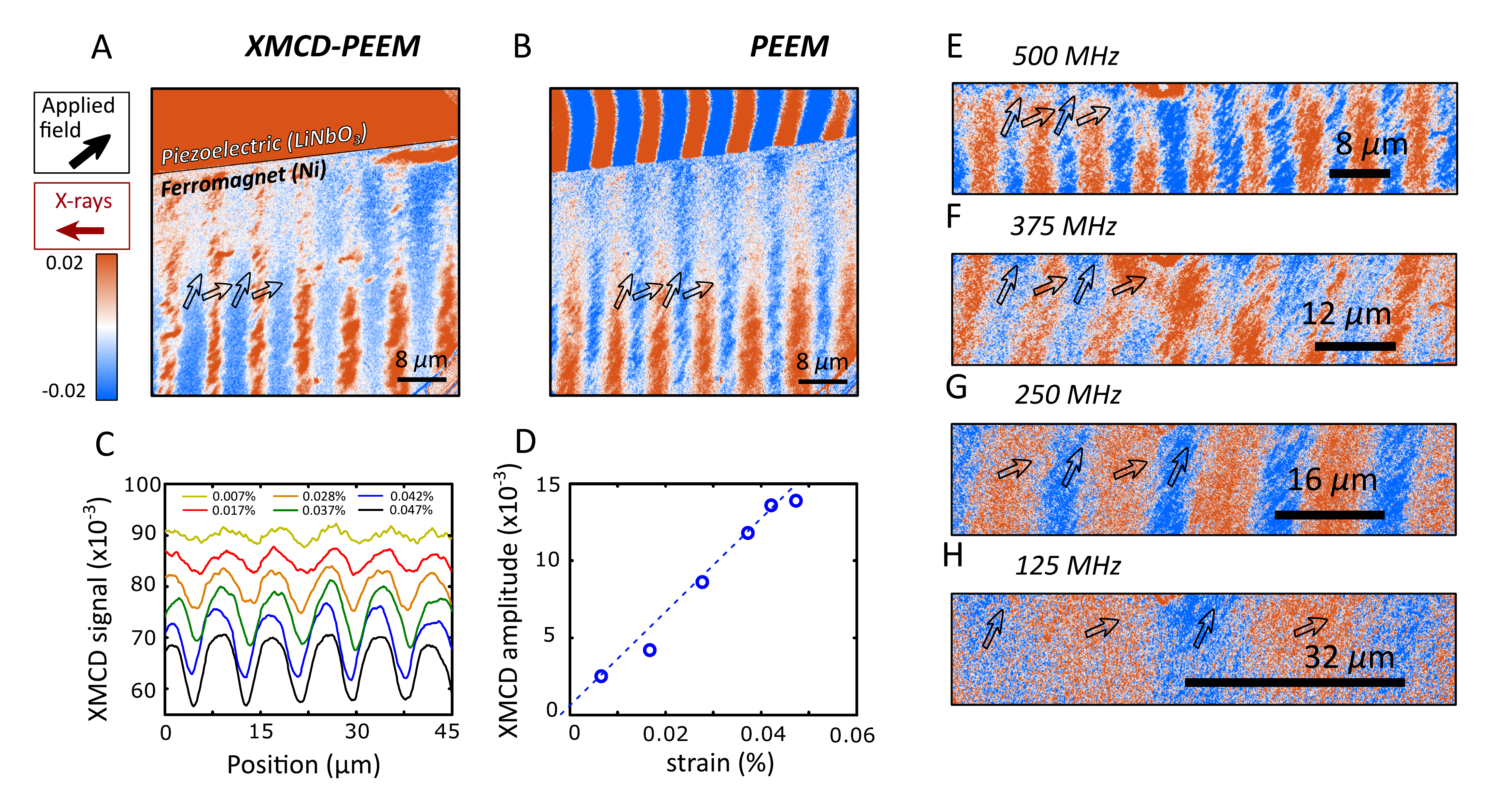}
	\caption{\small{\textbf{(A)} XMCD Image with a field of view of $50 \times 50\mu$m$^2$ showing both the ferromagnetic (Ni) and the piezoelectric (LiNbO$_3$) materials in presence of a SAW---stripes indicate the presence of a magnetic wave. \textbf{(B)} Image of the same location and conditions as in (A) obtained from the substraction of two PEEM images with opposite phases---stripes in the ferromagnet correspond to a magnetic wave and stripes in the piezoelectric correspond to the strain wave. \textbf{(C)} Profiles of the magnetoacoustic waves in the ferromagnet from XMCD images taken at different SAW amplitude. Each amplitude is used to calculate an associated strain value from the photoelectron shift of the local surface potential in the piezoelectric material. \textbf{(D)} Dependence of the magnetoacoustic waves with the SAW amplitude. \textbf{(E-H)} XMCD images of magnetoacoustic waves of frequencies 500, 375, 250, and 125 MHz having wavelengths of 8, 12, 16, and 32 $\mu$m.}}
	\label{fig2}
\end{figure}

\begin{figure}[htb!]
	\includegraphics[width=1\columnwidth]{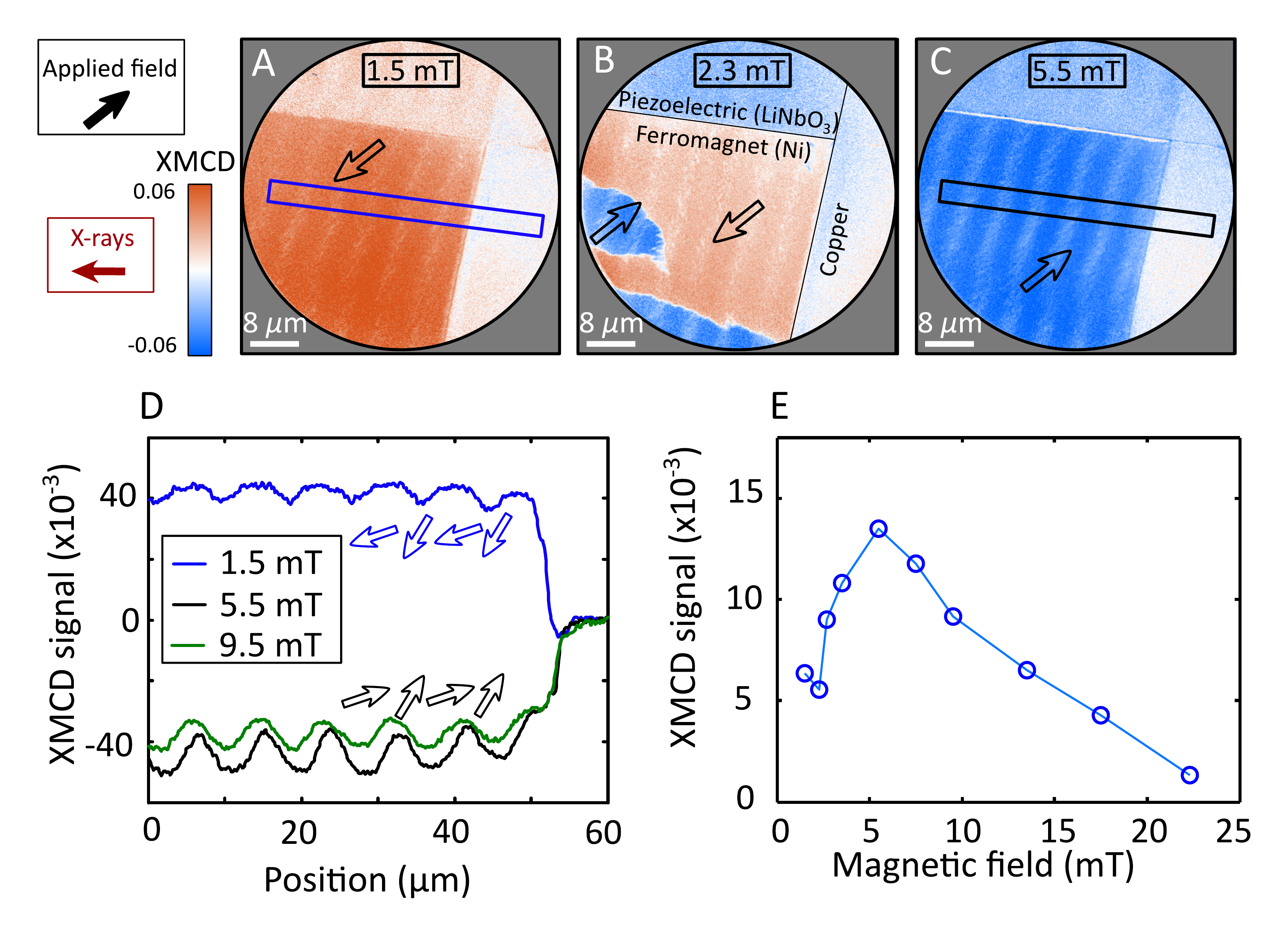}
	\caption{\small{\textbf{(A-C)} XMCD Image of a magnetoacoustic wave under different applied magnetic fields. The image includes the ferromagnetic (Ni) film and the piezoelectric (LiNbO$_3$) substrate together with a non-magnetic metallic structure (Cu). The arrows indicate the magnetization direction within the ferromagnetic film. \textbf{(D)} Horizontal cuts from \textbf{(A)} and \textbf{(C)} and from an additional field of 9.5 mT (image not shown). Schematic arrows indicate the magnetization direction at different wave points. \textbf{(E)} magnetoacoustic wave amplitude as a function of the external applied field.}}
	\label{fig3}
\end{figure}

\begin{figure}[htb!]
	\includegraphics[width=1\columnwidth]{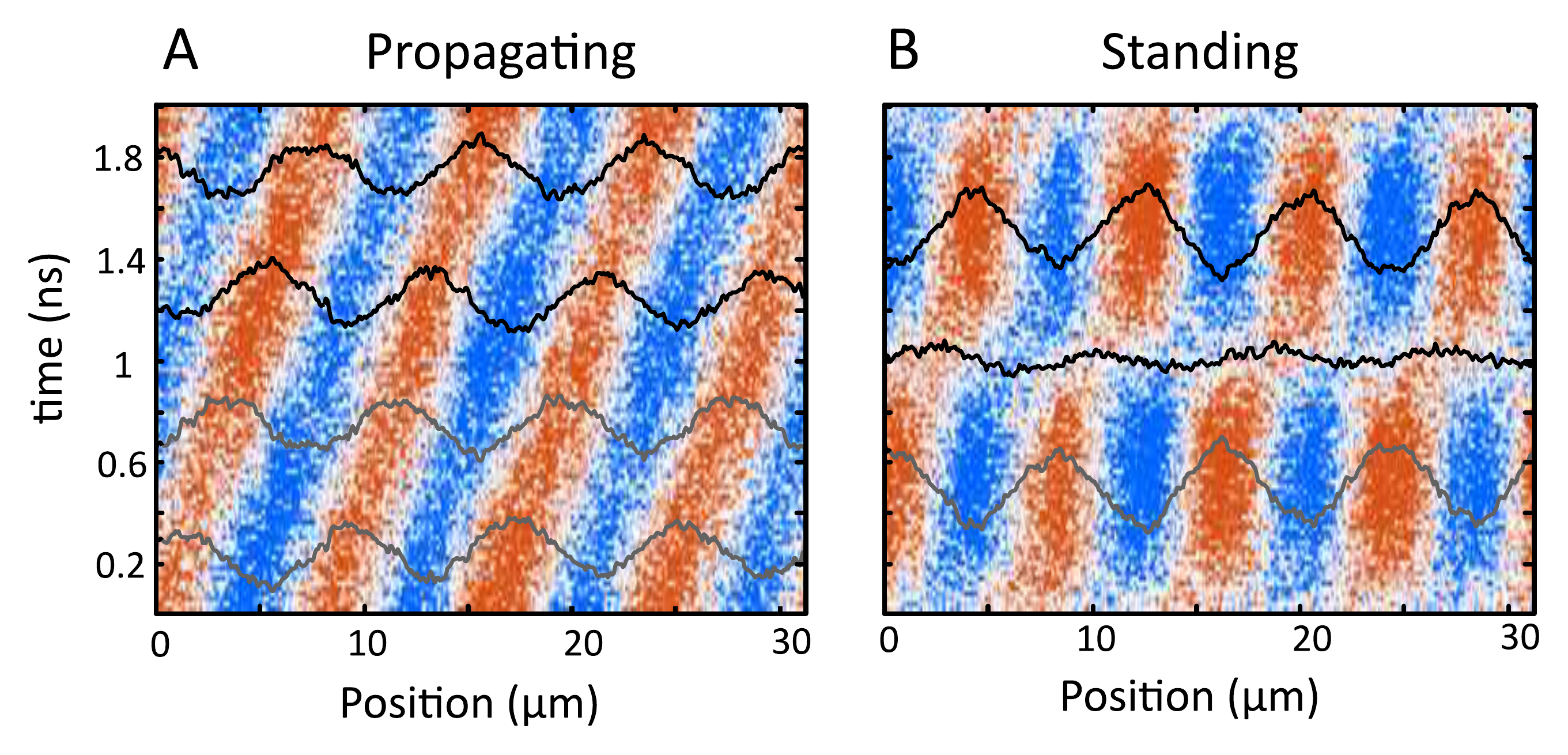}
	\caption{\small{Time evolution of a magnetoacoustic wave in a ferromagnetic material (Ni) for a propagating wave \textbf{(A)} and for a standing wave \textbf{(B)}. The waves are created with two IDT facing each other and separated 6 mm. A small detuning between the wave and the synchrotron frequencies combined with a subtraction of opposite phases allows for  the visualization of the magnetoacoustic waves. In A the maxima and minima are seen to shift position with time, while in B the positions are fixed but the amplitude oscillates periodically.}}
	\label{fig4}
\end{figure}

\end{document}